\documentclass{article}
\usepackage{spconf,amsmath,graphicx}
\usepackage[utf8]{inputenc} %
\usepackage[T1]{fontenc}    %
\usepackage{hyperref}       %
\usepackage{url}            %
\usepackage{booktabs}       %
\usepackage{amsfonts}       %
\usepackage{nicefrac}       %
\usepackage{microtype}      %
\usepackage{xcolor}         %
\usepackage{comment}
\usepackage{graphicx}
\usepackage{amsmath}
\usepackage{color}
\usepackage{appendix}
\usepackage{lipsum}
\usepackage{hanging}
\usepackage{wrapfig}
\usepackage{multirow}
\usepackage{enumitem}
\usepackage[skip=10pt]{subcaption}

\title{ProsoSpeech: Enhancing Prosody With Quantized Vector Pre-training in Text-to-Speech}
\name{Yi Ren$^{1}$, Ming Lei$^{2}$, Zhiying Huang$^{2}$,  Shiliang Zhang$^{2}$, Qian Chen$^{2}$, Zhijie Yan$^{2}$, Zhou Zhao$^{1}$}
\address{$^1$Zhejiang University, China, $^2$Speech Lab, Alibaba Group, China\\
rayeren@zju.edu.cn, \\ \{lm86501,zhiying.hzy,sly.zsl,tanqing.cq,zhijie.yzj\}@alibaba-inc.com, \\ zhaozhou@zju.edu.cn}
\begin{document}
\maketitle
\begin{abstract}
Expressive text-to-speech (TTS) has become a hot research topic recently, mainly focusing on modeling prosody in speech. Prosody modeling has several challenges: 1) the extracted pitch used in previous prosody modeling works have inevitable errors, which hurts the prosody modeling; 2) different attributes of prosody (e.g., pitch, duration and energy) are dependent on each other and produce the natural prosody together; and 3) due to high variability of prosody and the limited amount of high-quality data for TTS training, the distribution of prosody cannot be fully shaped. To tackle these issues, we propose ProsoSpeech, which enhances the prosody using quantized latent vectors pre-trained on large-scale unpaired and low-quality text and speech data. Specifically, we first introduce a word-level prosody encoder, which quantizes the low-frequency band of the speech and compresses prosody attributes in the latent prosody vector (LPV). Then we introduce an LPV predictor, which predicts LPV given word sequence. We pre-train the LPV predictor on large-scale text and low-quality speech data and fine-tune it on the high-quality TTS dataset. Finally, our model can generate expressive speech conditioned on the predicted LPV. Experimental results show that ProsoSpeech can generate speech with richer prosody compared with baseline methods.

\end{abstract}

\begin{figure*}[!t]
	\centering
	\begin{subfigure}[h]{0.32\textwidth}
		\centering
		\includegraphics[width=\textwidth,trim={0cm 0.05cm 11.2cm 0cm}, clip=true]{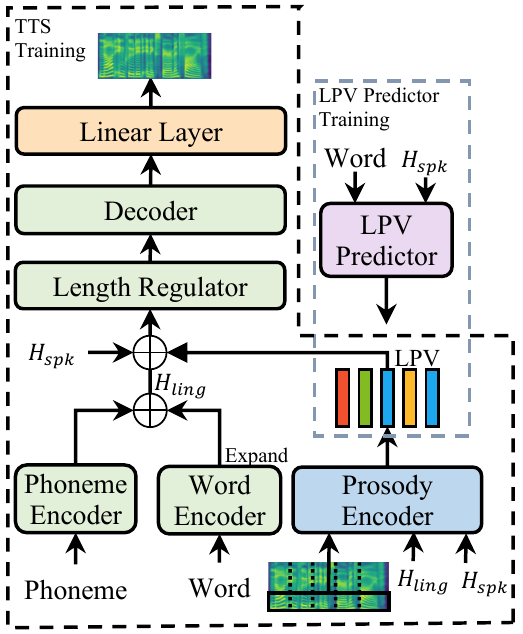}
		\vspace{-2mm}
		\caption{ProsoSpeech}
		\label{fig:arch_1}
	\end{subfigure}
	\begin{subfigure}[h]{0.22\textwidth}
		\centering
		\vspace{-4mm}
		\includegraphics[width=\textwidth,trim={5.25cm 0.00cm 7.9cm 0cm}, clip=true]{figs/arch.pdf}
		\vspace{-5mm}
		\caption{Prosody Encoder}
		\label{fig:arch_2}
	\end{subfigure}
	\begin{subfigure}[h]{0.215\textwidth}
		\centering
		\includegraphics[width=\textwidth,trim={8.7cm 0.0cm 4.1cm 0.00cm}, clip=true]{figs/arch.pdf}
		\vspace{0mm}
		\caption{LPV Predictor}
		\label{fig:arch_3}
	\end{subfigure}
	\begin{subfigure}[h]{0.19\textwidth}
		\centering
		\includegraphics[width=\textwidth,trim={12.5cm 0.0cm 1.0cm 0.0cm}, clip=true]{figs/arch.pdf}
		\vspace{-6mm}
		\caption{Pre-training}
		\label{fig:arch_4}
	\end{subfigure}
	\caption{The overall architecture for ProsoSpeech. 
	In subfigure (a), LPV denotes latent prosody vector; $H_{ling}$ and $H_{spk}$ denote the linguistic features and the speaker embedding respectively; 
	the "expand" operation after the word encoder is used to expand the word encoder outputs to make their length match that of phoneme encoder outputs. 
	In subfigure (b), 
	"Word-level Pooling" averages the hidden states within a word according to the word boundary.
	}
	\label{fig:arch}
	\vspace{-3mm}
\end{figure*}

\section{Introduction}
Recently, neural network-based text-to-speech (TTS) has been attracted a lot of attention~\cite{wang2017tacotron,shen2018natural,ping2018deep}. Thanks to the advance of non-autoregressive models~\cite{ren2019fastspeech,ren2020fastspeech} and powerful generative models~\cite{kim2020glow,lee2020multi,liu2021diffsinger,ren2021portaspeech}, the audio quality and inference speed of modern TTS has been greatly improved. However, synthesizing human-like expressive speech is still a challenging task.

To generate expressive speech, reference-encoder-based methods~\cite{stanton2018predicting,wang2018style,chien2021hierarchical,klapsas2021word} use autoencoder-like architecture to learn latent disentangled representations and successfully factorize speaker identity and prosody; prosody prediction-based methods~\cite{ren2020fastspeech,du2021mixture} first extract prosody attributes including pitch, duration and energy and use some modules to predict them conditioned on the input linguistic features. However, previous prosody modeling methods suffer from several issues: 1) some works use external tools to extract pitch contour. However, the extracted pitch has inevitable errors, such as v/uv decision errors and inaccurate F0 values. These errors not only degrade the performance of pitch prediction, but also hurt the optimization of TTS model, which is conditioned on the extracted pitch, and thus introduce some losses in prosody modeling. 2) Some works extract prosody attributes (e.g., pitch, duration and energy) from speech and model them separately. However, these prosody attributes are dependent on each other and produce natural prosody together. Modeling them separately may break their relationship and leads to unnatural prosody. 3) Prosody has very high variability and varies from person to person and word to word. It can be very difficult to shape the full distribution of prosody using the limited amount of high-quality TTS data.

To address these issues, in this paper, we propose ProsoSpeech, which enhances the prosody using quantized latent vectors pre-trained on the large-scale unpaired and low-quality text and speech data. Based on FastSpeech, our ProsoSpeech consists of the following designs: 
1) To avoid the errors in pitch extraction and take the dependency of prosody attributions into consideration, we introduce a word-level prosody encoder to disentangle the prosody from speech, which quantizes the low-frequency band of the speech to word-level quantized latent prosody vectors (LPV) according to the word boundary. To stabilize the training process of vector quantization and avoid index collapse, we further design a warm-up strategy with k-means cluster-based codebook initialization. 
2) Since we can extract prosody representation given any speech sample, we propose an autoregressive LPV predictor to predict LPV conditioned on word-level text sequence to model the prosody. 
3) To shape the distribution of prosody better, we pre-train the LPV predictor on large-scale text data and low-quality speech dataset and finetune it on the high-quality TTS dataset. Finally, we can generate expressive speech conditioned on the predicted LPV.

We conduct experiments on a high-quality Chinese TTS dataset and pre-train our LPV predictor on large-scale text corpus and low-quality speech datasets. The results show that ProsoSpeech can generate more natural speech with richer prosody and better audio quality compared with state-of-the-art TTS methods. We also conduct sufficient ablation studies to demonstrate the effectiveness of each design\footnote{We put some audio samples in \url{https://prosospeech.github.io/}.}.

\section{Our Method}
In this section, we introduce our proposed model ProsoSpeech. As shown in Figure \ref{fig:arch_1}, ProsoSpeech is based on FastSpeech and introduces some modules to further model the expressiveness of speech including the word encoder, the prosody encoder and the autoregressive latent prosody vector (LPV) predictor. In training, the input text sequence is converted to phoneme sequence and word sequence, which are encoded into linguistic features by phoneme and word encoder. Then the low-frequency part of the ground-truth mel-spectrogram is encoded to a quantized latent prosody vector (LPV) using the prosody encoder conditioned on the linguistic features. Finally, we feed the linguistic features and LPV together into the decoder to generate the predicted mel-spectrograms and optimize the model using mean square error (MSE) and similarity index measure (SSIM)~\cite{wang2004image} losses. So far, we disentangle the prosody from speech and obtain the prosody disentangled representation (LPV). To predict LPV sequence, we train an autoregressive LPV predictor conditioned on the word sequence. Besides, we employ a large-scale text and audio corpus to pre-train the LPV predictor to better understand the text context and shape the prosody distribution. In inference, since we do not have the ground-truth mel-spectrogram as a reference, we use the LPV predictor to predict LPV and generate expressive speech. In the following subsections, we introduce each module in detail.

\subsection{Prosody Encoder}
\label{sec:prosody_encoder}
The prosody encoder is designed to disentangle the prosody from speech using a word-level vector quantization bottleneck. As shown in Figure \ref{fig:arch_2}, the prosody encoder consists of two levels, each of them is a stack of convolution layers with ReLU activation and layer normalization. The first level compresses the mel-spectrograms into word-level hidden states according to the word boundary and the second level post-processes the word-level hidden states. Finally, these hidden states are fed into the EMA\footnote{the abbreviation of "exponential moving averages"}-based vector quantization layer~\cite{oord2017neural} to obtain the word-level LPV sequence. Since the timbre (speaker identity) and the content in speech are provided by speaker embedding and linguistic encoders (phoneme/word encoder) respectively, the LPV only contains the speaker-and-content-independent prosody information, due to the vector quantization bottleneck. Besides, we only take the low-frequency band of the mel-spectrogram (first 20 bins in each mel-spectrogram frame) as input to ease the difficulty of disentanglement, because it contains almost complete prosody and much less timbre/content information compared with the full band.

However, it takes thousands of steps of training for the prosody encoder to truly extract the prosody information from the word-level mel-spectrogram clips. So at the beginning of the training, the hidden states before vector quantization can be very noisy and meaningless. In this case, we find our prosody encoder tends to index collapse~\cite{rakhimov2020latent}, which means that some embedding vectors are close to a lot of encoders outputs and the model uses only a limited number of vectors from $e$. Index collapse severely limits the expression ability of our prosody encoder. To tackle this problem, we propose a warm-up strategy and k-means cluster-based centroid initialization: 1) we remove the vector quantization layer in the first 20k steps, making the prosody encoder extract the prosody information freely without any bottleneck; 2) after the first 20k steps, we initialize the codebook of the vector quantization layer with k-means cluster centers; 3) after initialization, we add the vector quantization layer as the prosody bottleneck for later training.

\subsection{Latent Prosody Vector Predictor}
Now that we have been able to extract the prosody representations using the prosody encoder, we can model the prosody by modeling the LPV sequence. As shown in Figure \ref{fig:arch_3}, LPV predictor is used to predict the word-level LPV sequence using text input, which adopts the self-attention-based ~\cite{vaswani2017attention} autoregressive architecture. Since the LPV sequence has the same length as the word sequence, we use the word-level context features as the condition, which is encoded by a context encoder in the LPV predictor. LPV predictor is trained in teacher forcing mode in the training stage and predicts the LPV autoregressively in inference.

\subsection{Pre-training and Fine-tuning}
Although the prosody representation can be modeled by the LPV predictor, it could be not accurate enough, due to the following reasons: 1) the text training data is not large enough (about 10k sentences) in TTS dataset, leading to poor context understanding for context encoder and difficulty in capturing the connection between the prosody and text. 2) The speech/prosody training data is not large enough, making the sample in prosody space somewhat sparse, resulting in inaccurate prosody distribution estimation. Thus, we propose a pre-training method using both additional pure text data and low-quality speech data as shown in Figure \ref{fig:arch_4}. For pre-training with text, the context encoder in LPV predictor is trained in a BERT-like~\cite{devlin2018bert} mask prediction manner with 0.15 masking probability. For pre-training with low-quality audio, the LPV predictor is pre-trained with the LPV sequence encoded from noisy audio. After these pre-training processes, we fine-tune the LPV predictor on the high-quality TTS dataset. Therefore, our final training pipeline includes TTS training (including the prosody encoder and main body of FastSpeech), pre-training context encoder with unpaired text, pre-training LPV predictor with low-quality speech and fine-tuning LPV predictor with high-quality TTS data in turn.

\section{Experiments}
\subsection{Experimental Setup} 
\textbf{Datasets} We evaluate ProsoSpeech on an internal Mandarin dataset, which contains 62,586 Mandarin audio clips (about 30 hours) and corresponding text transcripts. We split our TTS dataset into three subsets: 61,000 samples for training, 586 samples for validation and 1,000 samples for testing. We randomly choose 50 samples in the test set for subjective evaluation. We convert the text sequence to the phoneme sequence~\cite{arik2017deep,wang2017tacotron,shen2018natural,sun2019token,ren2019fastspeech} with an open-source Chinese grapheme-to-phoneme tool\footnote{\url{https://github.com/mozillazg/python-pinyin}}. We transform the raw waveform with the sampling rate 22050 into mel-spectrograms following \cite{shen2018natural,ren2019fastspeech} with the frame size 1024 and the hop size 256. For unpaired text data, we crawl 51M Chinese sentences from the internet. For low-quality audio data, we use an internal Chinese ASR dataset which contains about 300 hours of audio clips. The speaker embeddings are extracted using resemblyzer\footnote{\url{https://github.com/resemble-ai/Resemblyzer}}.

\noindent\textbf{Model Configuration} The phoneme encoder, word encoder, and decoder all adopt the Transformer proposed in FastSpeech~\cite{ren2019fastspeech}, whose number of layers, hidden size, kernel size and the filter size are set to 4, 192, 5 and 384 respectively. The size of the pre-trained speaker embeddings are projected to 192 using a dense layer. Each convolution stack in prosody encoder contains 5 layers of 1D Convolution, ReLU and layer normalization. The default size of the codebook in the vector quantization layer is set to 128. The autoregressive LPV predictor contains 3 Transformer layers and the context encoder contains 6, with the hidden size 384, kernel size 5 and the filter size 384. The context encoder are shared among all LPV predictor layers. 

\noindent\textbf{Evaluation} The output mel-spectrograms of our model are transformed into audio samples using HiFi-GAN~\cite{kong2020hifi}\footnote{\url{https://github.com/jik876/hifi-gan}} trained in advance. We conduct the MOS (mean opinion score) evaluation on the test set to measure the audio quality. We keep the text content consistent among different models to exclude other interference factors, only examining the audio quality or prosody. Each audio is listened by at least 20 testers, who are all native English speakers. They are told to listen all samples in the quiet indoor environment with earphones and the volume should be at least 75\% to make them distinguish the details of all audio samples. We also use average pitch dynamic time warping (DTW)~\cite{muller2007dynamic} distance ($D_{pit}$) and duration KL-divergence ($\text{KL}_{dur}$) as the objective metrics to evaluate the overall prosody: average pitch DTW distance is defined as $D_{pit}=\text{DWT}(p_1, p_2)/l_{path}$, where $p_1$ and $p_2$ are two pitch contours (only voiced part is considered), $\text{DWT}(\cdot)$ calculates the minimal DTW distance and $l_{path}$ is the length of best DTW path; duration KL-divergence is defined as $\text{KL}_{dur}=\sum_{w \in \mathcal{W}}^{}\text{KL}(\text{KDE}(d^w_1), \text{KDE}(d^w_2)) / \text{S}_\mathcal{W}$, where $\mathcal{W}$ is the Chinese character dictionary, $\text{KL}(\cdot)$ calculates the KL-divergence, $\text{KDE}(\cdot)$ is the kernel density estimation~\cite{terrell1992variable} function smoothing the discrete distribution and converting it to continuous distribution, $d^w_i$ is the duration discrete distribution of word $w$ from system $i$ and $\text{S}_\mathcal{W}$ is the size of the dictionary $\mathcal{W}$. The duration values and pitch contours for measurement are extracted using external tools.

\begin{table}[!t]
\small
\caption{The audio performance (MOS), pitch accuracy ($D_{pit}$), duration accuracy ($\text{KL}_{dur}$) comparisons. Best results are marked in bold.}
\centering
\begin{tabular}{ l | c | c | c }
\toprule
Method &  MOS $\uparrow$ & $D_{pit}$ $\downarrow$ & $\text{KL}_{dur}$ $\downarrow$ \\
\midrule
\textit{GT}                                    & 4.39 $\pm$ 0.09 & /  & / \\
\textit{GT (voc.)}                             & 4.08 $\pm$ 0.08 & /  & / \\
\midrule 
\textit{FastSpeech~\cite{ren2019fastspeech}}   & 3.65 $\pm$ 0.12 & 11.71 & 0.136 \\
\textit{FastSpeech 2~\cite{ren2020fastspeech}} & 3.78 $\pm$ 0.11 & 11.79 & 0.143 \\
\textit{FastSpeech 2 (joint)}                  & 3.42 $\pm$ 0.13 & 12.21 & 0.145 \\
\midrule
\textit{ProsoSpeech}                           & \textbf{3.85 $\pm$ 0.09} & \textbf{10.26} & \textbf{0.131} \\
\bottomrule
\end{tabular}
\label{tab:main_results}
\vspace{-3mm}
\end{table}

\subsection{Performance}
We compare the audio quality and prosody of generated audio samples of our ProsoSpeech with other systems, including 1) \textit{GT}, the ground truth audio; 2) \textit{GT (voc.)}, where we first convert the ground truth audio into mel-spectrograms, and then convert the mel-spectrograms back to audio using HiFi-GAN; 3) \textit{FastSpeech~\cite{ren2019fastspeech}}; 4) \textit{FastSpeech 2~\cite{ren2020fastspeech}} and 5) \textit{FastSpeech 2 (joint)}: train FastSpeech 2 with joint TTS and ASR dataset. The results are shown in Table \ref{tab:main_results}. We have the following observations: 1) For audio quality, ProsoSpeech outperforms previous TTS models in terms of MOS, which shows the superiority of our proposed method. 2) For pitch accuracy, compared with previous methods, $D_{pit}$ of generated audio of \textit{ProsoSpeech} is smaller than that of other methods, demonstrating the powerful and efficient prosody modeling ability of ProsoSpeech. 3) For duration accuracy, compared with previous methods, the duration predicted by \textit{ProsoSpeech} is more close to the ground-truth duration in distribution-level according to $\text{KL}_{dur}$, indicating that ProsoSpeech can model the duration distribution better. 4) Compared with \textit{FastSpeech 2 (joint)}, ProsoSpeech also achieves better performance, which demonstrates that the prosody extraction in our prosody encoder is necessary for latter low-quality data pre-training. We also find the performance of \textit{FastSpeech 2 (joint)} is worse than that of \textit{FastSpeech 2}, which shows that if we directly pre-train the TTS model with additional low-quality speech, it can disturb the model training and even results in worse quality.

\subsection{Ablation Study and Analyses}
We conduct ablation studies to demonstrate the effectiveness of designs in ProsoSpeech, including the k-means initialization, text pre-training and audio pre-training, and also explore the appropriate size of codebook in prosody encoder. We conduct CMOS evaluation for these ablation studies. The results are shown in Table \ref{tab:abl}. From the table, we can see that: 1) from Line 2 to 4, both text and speech pre-training can improve the pitch and duration accuracy, and they can work together to further improve the prosody; 2) from Line 5, k-means initialization in prosody encoder can improve the prosody and stabilize the training; 3) from Line 6 and 7, the default size of the codebook (128) is enough for the prosody modeling in our setting, and the performance can be degraded if we use a smaller codebook.

\begin{table}[!t]
\small
\caption{The audio performance (CMOS), pitch accuracy ($D_{pit}$), duration accuracy ($\text{KL}_{dur}$) comparisons for ablation studies. "k-means init." denotes k-means initialization in the prosody encoder. "PT" denotes pre-training. }
\centering
\begin{tabular}{ l | c | c | c }
\toprule
Method &  CMOS & $D_{pit}$ $\downarrow$ & $\text{KL}_{dur}$ $\downarrow$ \\
\midrule
\textit{ProsoSpeech}              & 0.00  & \textbf{10.26} & 0.131 \\
\midrule 
~\textit{w/o text PT}             & -0.22  & 10.65 & 0.135 \\
~\textit{w/o audio PT}            & -0.17  & 10.51 & 0.136 \\
~\textit{w/o text/audio PT}       & -0.25  & 10.70 & 0.136 \\
~\textit{w/o k-means init.}       & -0.31  & 11.52 & 0.141 \\
~\textit{w/o size of $e=64$}      & -0.14  & 10.32 & 0.131 \\
~\textit{w/o size of $e=256$}     & -0.03  & 10.28 & \textbf{0.129} \\

\bottomrule
\end{tabular}
\label{tab:abl}
\vspace{-3mm}
\end{table}

\section{Conclusion}
In this paper, we proposed ProsoSpeech, enhancing the prosody using quantized latent vectors which are pre-trained on large-scale text and noisy speech data. To model all attributes of prosody together, we first introduced a word-level prosody encoder, which quantizes the word-level prosody of the speech and compresses all prosody attributes in the quantized latent prosody vectors (LPV). Then we proposed an LPV predictor to predict LPV given word-level text sequence. To better shape the complex distribution of prosody space, we pre-train the LPV predictor on large-scale text data and low-quality speech dataset and finetune it on a high-quality TTS dataset. Our experimental results demonstrated the effectiveness of our method and each design. In the future, we will introduce more powerful generative models to improve the audio quality of speech. We can also leverage longer context information such as paragraph-level context or dialogue-level context to further improve the prosody of speech.

\section{Acknowledgments}
This work was supported by Alibaba Group through Alibaba Innovative Research Program.

\bibliographystyle{IEEEbib}
\bibliography{refs}

\end{document}